\documentclass[aps, prd, preprintnumbers, longbibliography, superscriptaddress, nofootinbib, floatfix, notitlepage]{revtex4-1}

\usepackage{amsmath}
\usepackage{bm}
\usepackage{amsfonts}
\usepackage{graphicx}
\usepackage{hyperref}
\usepackage[dvipsnames]{xcolor}
\usepackage[normalem]{ulem}
\usepackage{xspace}
\usepackage{relsize}
\usepackage{aas_macros}
\newcommand{\primat}{\texttt{PRIMAT}\xspace}
\newcommand{\parthenope}{\texttt{PArthENoPE}\xspace}

\newcommand{\Neff}{N_\mathrm{eff}}

\newcommand{\Yp}{Y_\mathrm{p}}
\renewcommand{\DH}{\mathrm{D/H}}
\newcommand{\dd}{\mathrm{d}}
\newcommand{\obs}{\mathrm{obs}}

\hypersetup{
colorlinks=true,
linkcolor=blue,
linktoc=page,
citecolor=blue,
urlcolor=blue}

\hypersetup{colorlinks=true,
linkcolor=Mulberry,
citecolor=NavyBlue,
filecolor=Mulberry,
urlcolor=NavyBlue,
menucolor=BrickRed,
runcolor=Mulberry
}

\begin{document}
\preprint{N3AS-25-006}

\title{Primordial neutrinos fade to gray: constraints from cosmological observables}

\newcommand{\FIRSTAFF}{\affiliation{Department of Physics, University of California Berkeley, Berkeley, CA 94720, USA}
\affiliation{Department of Physics, University of California San Diego, La Jolla, CA 92093, USA}}
\newcommand{\SECONDAFF}{\affiliation{Departament de Física Teórica and IFIC, Universitat de València-CSIC, E-46100, Burjassot, Spain}}
\newcommand{\THIRDAFF}{\affiliation{Institut d’Astrophysique de Paris, CNRS UMR 7095, Sorbonne Université, 98 bis Bd Arago, 75014 Paris, France}}

\author{Gabriela Barenboim}
\email[Electronic address: ]{gabriela.barenboim@uv.es}
\SECONDAFF
\author{Julien Froustey}
\email[Electronic address: ]{jfroustey@berkeley.edu}
\FIRSTAFF
\author{Cyril Pitrou}
\email[Electronic address: ]{cyril.pitrou@iap.fr}
\THIRDAFF
\author{Héctor Sanchis}
\email[Electronic address: ]{hector.sanchis@uv.es}
\SECONDAFF

\date{\today}
\begin{abstract}
We investigate the effect of potentially large distortions of the relic neutrino spectra on cosmological observables. To that end, we consider a phenomenological model of “gray” spectral distributions, described by a single parameter, which generalizes the traditional $y$-distortions to possibly large negative values. Implementing these distortions in the primordial nucleosynthesis code \primat, we can constrain the distortion parameter along with the presence of extra radiation, exploiting the complementarity of big bang nucleosynthesis and cosmic microwave background measurements to disentangle gravitational and non-thermal effects. These constraints rule out a distortion where more than $\sim 1/2$ of the neutrinos' energy density is replaced by dark radiation. Nonetheless, we find that large distortions, accompanied with extra radiation, are allowed—and even slightly preferred in some cases—by current cosmological observations. As this scenario would require substantial modifications to the physics of neutrino decoupling in the early Universe, these observational constraints call for a renewed attention on the possibility of large deviations from the standard cosmological model in the neutrino sector.
\end{abstract}

\maketitle

\section{Introduction}

Neutrinos play an important role in our current understanding of cosmology. In the standard hot big bang model, they were initially in thermal equilibrium due to weak interactions, and eventually decoupled slightly before the onset of Big Bang Nucleosynthesis (BBN), forming a cosmic neutrino background whose spectrum is expected to be nearly thermal.

The effect of neutrinos can be seen in various cosmological observables~\cite{NeutrinoCosmology}, such as the Cosmic Microwave Background (CMB). It is primarily sensitive to the energy density of the cosmic neutrino background, allowing one to measure the effective number of relativistic species $\Neff$, consistently with the standard prediction of $3.044$~\cite{Froustey:2020mcq,Bennett:2020zkv,Drewes:2024wbw}. A complementary probe is given by spectroscopic measurements of the primordial abundances, since neutrinos set the conditions under which BBN takes place~\cite{Grohs:2023voo}. The agreement between theory (with publicly available codes like \primat~\cite{Pitrou:2018cgg}, \parthenope~\cite{Gariazzo:2021iiu} or \texttt{PRyMordial}~\cite{Burns:2023sgx}) and observations for deuterium and helium abundances is a major success of the standard cosmological model, slightly obscured by the so-called lithium problem~\cite{Fields:2011zzb}. While these two probes are sensitive to the epoch when neutrinos were ultrarelativistic, the growth of large-scale structures is affected by the neutrino masses, which are known to be nonzero because of the phenomenon of neutrino oscillations~\cite{Kajita:2016cak,McDonald:2016ixn}. However, this cosmological effect has not been seen yet. Quite the opposite — the recent results by DESI~\cite{DESI:2024mwx,DESI:2025zgx} do not show preference for massive neutrinos, and if one allows for negative effective masses, there is a $\sim 3 \sigma$ tension between the cosmological measurements and the minimum value compatible with neutrino flavor oscillations~\cite{2024JHEP...09..097C_negativenumass,Jiang:2024viw}.

This growing tension calls for the study of non-standard features in the cosmic neutrino spectrum. Moreover, although early Universe measurements (CMB, BBN) are consistent with the standard scenario of neutrino decoupling, it is worth emphasizing that they are only indirect probes of the early cosmic history. Specifically, highly non-standard scenarios that are still compatible with measurements simply cannot be excluded~\cite{Cuoco:2005qr,Alvey:2021sji}. In this article, we thus focus on non-standard neutrino spectral distortions.

In the standard scenario, percent-level non-thermal distortions are present because of the out-of-equilibrium decoupling process~\cite{Dolgov:1997mb,Mangano:2005cc,Grohs:2015tfy}, but their effects are small on cosmological observables (see e.g.,~\cite{Froustey:2019owm}).  A popular scenario assumes nonzero neutrino/antineutrino asymmetries, which can be associated to nonzero chemical potentials. Such a change modifies in particular the neutron-to-proton ratio at the onset of BBN, allowing one to constrain these asymmetries~\cite{Simha:2008mt,Shimon:2010ug,Castorina:2012md,Barenboim:2016shh,Oldengott:2017tzj,Escudero:2022okz,Froustey:2024mgf,Lattanzi:2024hnq,Domcke:2025lzg}. Here, we consider potentially large deviations from the standard Fermi-Dirac spectrum, identical for neutrinos and antineutrinos, starting from $y$-type distortions similar to the ones introduced for the neutrino spectrum in~\cite{Barenboim:2024wek}. We generalize them to describe both an increase or decrease of the neutrino energy density, and allow for the presence of additional dark radiation. Specifically, we consider a so-called “gray” distribution, a type of distribution usually discussed in the context of CMB (see, e.g.,~\cite{Stebbins:2007ve,Pitrou:2014ota}).

This paper is structured as follows: in Sec.~\ref{SpectralDistortionsIntro}, we introduce our model for a distorted cosmic neutrino spectrum, and study its theoretical implications for cosmological observables, with special emphasis on BBN. In Sec.~\ref{BBNTheoryCalculations}, we employ the BBN code \primat to quantitatively describe the effects of this distorted spectrum. Then, in Sec.~\ref{ConstraintsFromObservations}, we present constraints on our model from cosmological observations. Finally, we discuss the implications of those constraints and conclude in Sec.~\ref{sec:Conclusions}.
Throughout this paper, we use natural units where $\hbar = c = k_B = 1$.

%%%%%%%%%%%%%%%%%
\section{Spectral Distortions in the Neutrino Sector}
\label{SpectralDistortionsIntro}
%%%%%%%%%%%%%%%%%

%%%%%%
\subsection{Theory and model}
%%%%%%

\subsubsection{$y$ and $\mu$-type distortions}
\label{subsubsec:ymudist}

In the CMB, nonthermal features in the distribution function of background photons can be divided into two distinct types: $\mu$-type and $y$-type distortions. These distortions, which are well known, well studied, and bounded to be very small in the CMB spectrum, can arise from inefficient (efficient) Compton scattering with electrons for the $y$ ($\mu$) type.
In the case of the cosmic neutrino background, the same types of distortions can be used. 
Furthermore, $\mu$-type distortions can be attributed to the chemical potentials of (anti)neutrinos, which emerge if the lepton asymmetry of the Universe is significant, a possibility allowed by current experimental data~\cite{Simha:2008mt,Shimon:2010ug,Castorina:2012md,Barenboim:2016shh,Oldengott:2017tzj,Escudero:2022okz,Froustey:2024mgf,Lattanzi:2024hnq,Domcke:2025lzg}. In this case, however, these “distortions” should not be interpreted as such, since a chemical potential is a thermal albeit non-standard feature of the neutrino distribution function. We also note that low reheating temperature scenarios can lead to $\mu$-type-like distortions, and the impact on cosmological observables allows one, in turn, to constrain these scenarios in the same spirit as this work~\cite{Kawasaki:1999na,Kawasaki:2000en,deSalas:2015glj,Hasegawa:2019jsa,Barbieri:2025moq}.

For $y$-type distortions, on the other hand, no obvious realization within the neutrino sector is evident. In analogy to the Sunyaev-Zeldovich effect \cite{1969Ap&SS...4..301Z_Sunyaev_Zeldovich}, such a distortion could in principle be produced by neutrinos scattering on some high-energetic particles (in a dark sector) after neutrino decoupling, such that the associated $y$ parameter (that we will call $y_\mathrm{SZ}$) would quantify the amount of energy transferred to the neutrinos by this hot gas. As a consequence, the spectrum of neutrinos would no longer be a perfect Fermi-Dirac distribution and would be characterized by a decrease in intensity at lower frequencies and an increase at higher frequencies.
A distinguishing feature of a $y$-type distortion is that the number of relic particles remains constant, while the energy injected is proportional to $y_\mathrm{SZ}$. Such a distorted Fermi-Dirac spectrum reads
\begin{equation}
\label{eq:f_y_SZ}
\begin{aligned}
  f_\nu(x,y_\mathrm{SZ}) &\equiv \hat{f}_\nu(x)+ y_\mathrm{SZ} \frac{1}{x^2} \frac{\dd}{\dd x} \left[x^4 \frac{\dd}{\dd x}\hat{f}_\nu(x) \right] \\
  &=\hat{f}_\nu(x)  \left[1+ y_\mathrm{SZ} \frac{e^x x}{e^x+1} \left(x\frac{e^x-1}{e^x+1}-4\right)\right] \, , 
\end{aligned}
\end{equation}
where the Fermi-Dirac distribution is
\begin{equation}
\label{DefFD}
\hat{f}_\nu(x) \equiv \frac{1}{e^x + 1}\,,
\end{equation}
and with $x=E/T$ the dimensionless ratio of the neutrino energy $E$ and the reference temperature $T$. For a given neutrino distribution, the number density and the energy density are obtained via integration in phase space as
\begin{equation}
n_\nu = \frac{T^3}{2 \pi^2} \int_0^\infty f_\nu(x) \, x^2 \, \mathrm{d} x\,,\qquad \rho_\nu = \frac{T^4}{2 \pi^2} \int_0^\infty f_\nu(x) \,  x^3 \, \mathrm{d} x\,.
\end{equation}
Hence the variation of energy density induced by the distortion in~\eqref{eq:f_y_SZ} is given by
\begin{equation}
    \frac{\delta \rho_\nu}{\hat\rho_\nu} = 4 y_\mathrm{SZ} \, ,
\end{equation}
where $\hat{\rho}_\nu$ (resp. $\hat{\rho}_\nu + \delta \rho_\nu$) is the energy density calculated with $\hat{f}_\nu$ [resp. $f_\nu(x,y_\mathrm{SZ})$].
Another little-known feature is that the $y_\mathrm{SZ}$ parameter must be positive. This limitation can be understood by seeing the spectrum as a sum of 
blackbodies\footnote{Although for photons the “blackbody” spectrum is the Planck distribution, we can also use this term for neutrinos, implying a Fermi-Dirac distribution.} with different temperatures, and characterizing the distribution of the temperatures by its moments in temperature space.
Stebbins~\cite{Stebbins:2007ve} has shown that the superposition of blackbodies is better characterized as a distribution in the variable $\log(T)$ and not $T$.
By doing so, it can be easily demonstrated~\cite{Pitrou:2014ota} that the average of $\log(T)$ corresponds to the average of the temperatures, while the variance of the distribution, obviously a positive quantity, can be mapped directly to what is usually called a $y_\mathrm{SZ}$-distortion. More directly, one can also see that the function~\eqref{eq:f_y_SZ} is asymptotically equivalent, for $x \to \infty$, to $y_\mathrm{SZ} x^2 e^{-x}$, which has to be positive for $f_\nu$ to be a distribution function.
Extensions and higher order corrections, normally called down-Comptonization (see, e.g.,~\cite{Miao:2021vuj}), do exist in the literature, however they break down at relatively low energy depletion.

\subsubsection{Generalized distortions}

In the case of neutrinos, where we want to explore significant deviations from a thermal spectrum, a new type of distortion is therefore needed, which mimics the “standard” $y_\mathrm{SZ}$-distortion in the positive parameter regime and can be safely extended to the negative one. Namely, we want to define a distortion which can describe an injection or depletion of energy, proportional to some “$y$” parameter, while keeping the number of particles constant. 

A good starting point is to consider again the resulting spectrum as a weighted sum of Fermi-Dirac spectra with different temperatures, 
\begin{equation}\label{Eqffromp}
 f_\nu(x)= \int_0^\infty p(t) \,  \hat{f}_\nu\left( \frac{x}{t} \right) \mathrm{d}t \, ,
\end{equation}
where $p(t)$ is a distribution function of temperatures $t$ (in units of the reference temperature $T$). If $\int_0^\infty  p(t) \, \mathrm{d}t\neq 1$ the distribution is called “gray,” and the grayness parameter~\cite{Stebbins:2007ve} is defined by $1-g \equiv \int_0^\infty  p(t) \, \mathrm{d}t$.
A full description of gray distributions and a complete generalization of the $y_\mathrm{SZ}$ distortion is beyond the scope of this work and will be given elsewhere.

In this framework, a distribution of relative temperatures, which preserves the number density while modifying the energy density, can be achieved for $y>0$ with the distribution
\begin{equation}
p_y(t) = \begin{cases}
		\dfrac{1}{2 |y|}t^{-3} & \text{if $1\leq t\leq 1+ 2 y $}\,,\\
            0 & \text{otherwise}.
		 \end{cases}\, 
\end{equation}
When $y<0$, the same functional form is taken but in the range $1+2y \leq t\leq 1$. The energy density of the distribution resulting from~\eqref{Eqffromp} is then related to the Fermi-Dirac one through $\rho_\nu = \hat\rho_\nu(1+y)$, whereas $n_\nu = \hat{n}_\nu$.
An important feature is that this distribution is gray since $1-g = (1+y)/(1+2y)^2$, and this relates to the possibility of having an energy density variation either positive or negative depending on the sign of $y$. 

\begin{figure}[!h]
\centering
\includegraphics[width=0.95\textwidth]{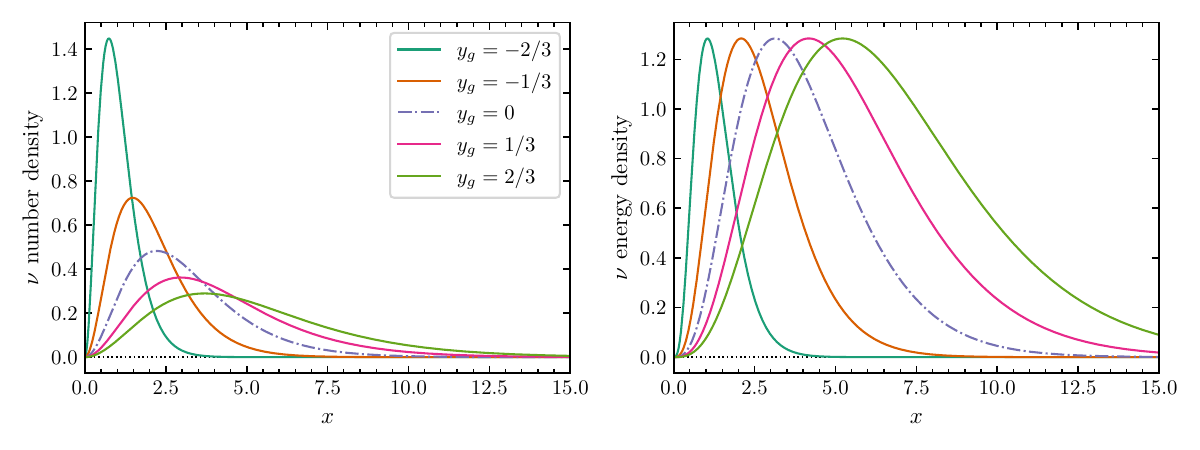}
\caption{Illustration of the effect of the distortion~\eqref{eq:normalization_g} on the neutrino differential number density [$x^2 f_\nu(x,y_g)$, left], and the differential energy density [$x^3 f_\nu(x,y_g)$, right].}
\label{fig:ydistortions}
\end{figure}

For the sake of this work, this motivates us to consider the even simpler gray distribution\footnote{We will assume for simplicity that all neutrino species have the same distribution function. This is somehow motivated by the action of neutrino flavor oscillations at MeV temperatures, although a specific model should be carefully studied.}
\begin{equation}
p_{y_g}(t) \equiv \frac{1}{(1+y_g)^3} \, \delta\left(1+y_g - t\right)\,,
\end{equation}
where the parameter $y_g$ must be larger than $-1$. The grayness of this distribution is $g = 1- (1+y_g)^{-3}$, and with~\eqref{Eqffromp} this corresponds to the simple spectrum
\begin{equation}
\label{eq:normalization_g}
 f_\nu\left( x,y_g \right) \equiv \frac{1}{\left(1+y_g\right)^3} \;  \hat{f}_\nu\left( \frac{x}{y_g+1}\right) \,.
\end{equation}
This distribution function never becomes negative while the neutrino number density remains equal to the one of the Fermi-Dirac distribution~\eqref{DefFD}\footnote{One can directly see that $\int_{0}^{\infty}{\dd{x} \, x^2 f_\nu(x,y_g)} = \int_{0}^{\infty}{\dd{x} \, x^2 \hat{f}_\nu(x)}$ via the change of variables $x' = x/(y_g+1)$.} and the energy injection (or depletion)
is 
\begin{equation}
\label{eq:rho_yg}
    \frac{\delta \rho_\nu}{\hat{\rho}_\nu} = y_g \, .
\end{equation}
We can then consider this gray distribution as a generalized $y$-type distortion, with a physical behavior for $y_g \leq 0$. We represent the differential number and energy densities associated to the distribution~\eqref{eq:normalization_g} on Fig.~\ref{fig:ydistortions}.

\begin{figure}[!ht]
\centering
\includegraphics[width=0.95\textwidth]{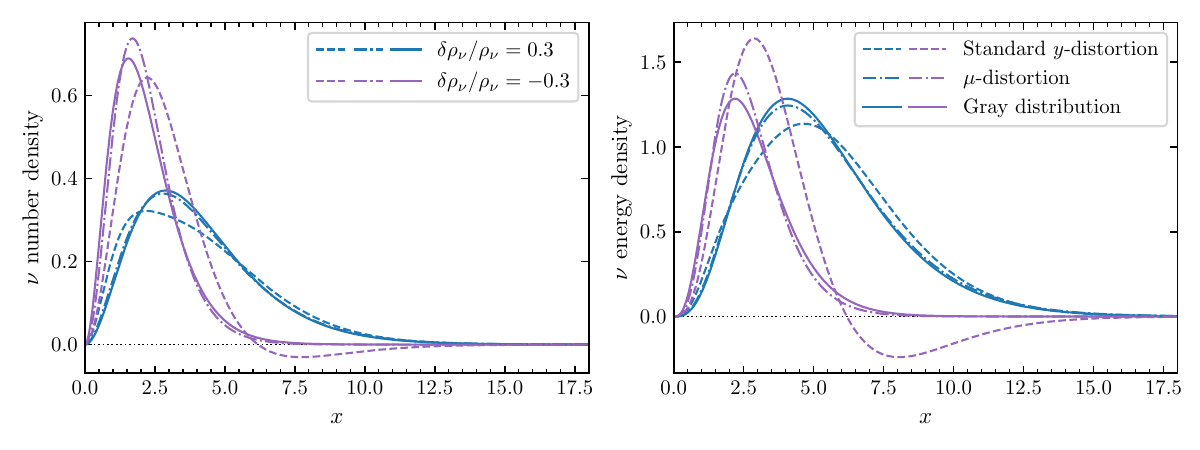}
\caption{Comparison of the standard $y$-distortion [Eq.~\eqref{eq:f_y_SZ}, dashed], the $\mu$-distortion [see text, dash-dotted] and the gray distribution [Eq.~\eqref{eq:normalization_g}, solid] for the same energy injection/depletion. The two panels are identical to Fig.~\ref{fig:ydistortions}.}
\label{fig:ydistortions_comparison}
\end{figure}

Although the distortion we consider is phenomenological and allows us to probe generally non-standard neutrino features, it is worth noting that it is not physically unmotivated. If this distortion is created by \emph{number-conserving} (elastic scattering) interactions of neutrinos with a dark sector at a different temperature, then sufficiently efficient interactions will cause the spectrum to be thermal with a chemical potential so as to preserve the number density while allowing a modified energy density. In addition, under this hypothesis of neutrinos and antineutrinos number densities being separately conserved, their respective chemical potentials need not be opposite as is the case when annihilation reactions (with $e^+e^-$, which have negligible degeneracy) take place. In Fig.~\ref{fig:ydistortions_comparison} we compare, for a positive and negative energy density variation at constant number density, the associated $y_\mathrm{SZ}$-distortion \eqref{eq:f_y_SZ}, the thermal distribution with chemical potential satisfying these conditions,\footnote{The unique distributions $\hat{f}_\nu\left[(x-\mu/T)/t\right]$ with chemical potential $\mu$ and temperature $T' = t T$ such that $n_\nu = \hat{n}_\nu$ and $\delta \rho_\nu/\hat{\rho}_\nu = + 0.3$ ($ - 0.3$) are such that $t \simeq 1.336$ ($0.633$) and $\mu/T \simeq - 1.237$ ($1.022$).} and our gray distribution \eqref{eq:normalization_g}. As discussed in Sec.~\ref{subsubsec:ymudist}, the standard $y_\mathrm{SZ}$-distortion is unphysical in the negative $y_\mathrm{SZ}$ regime. The other two distributions are physical in all cases, and we can see that the gray spectrum distortion is rather similar to the one associated with a chemical potential. The gray distortion is more convenient from a theoretical point of view since we can directly connect the energy density variation to a single parameter, $y_g$, instead of solving for a new temperature and chemical potential. More broadly, gray distortions in general are also physically motivated. A more direct possibility to generate a gray spectrum with $g>0$ is through the disappearance of neutrinos to a dark sector, regardless of their energy, as has already been considered but severely constrained for photons~\cite{Ellis:2013cu}. Such a distortion would not conserve the number of neutrinos, and could thus have interesting consequences for the late-time neutrino behavior. Thus, gray distortions other than the one considered here could also exist and be generated by other types of reactions (e.g., creation/annihilation reactions).

\subsubsection{Distortions and incomplete neutrino decoupling} 

Spectral distortions cannot survive if they are created prior to neutrino decoupling. Indeed, the strong neutrino self-interactions and interactions with $e^\pm$ (scattering, pair creation/annihilation) would force the thermalization of neutrino spectra. In the standard cosmological scenario, neutrinos decouple at a temperature $T_\mathrm{dec} \sim 1.3 - 1.5 \, \mathrm{MeV}$~\cite{Dolgov:2002wy}. As we discuss in detail later, the main effect of spectral distortions is to modify the freeze-out of neutron/proton interconversions prior to BBN, which occurs around $T_\mathrm{FO} \sim 0.8 \, \mathrm{MeV}$. There is thus a short window where non-standard physics should be operating in order to create the distortions we are considering in this work. However, the very existence of non-standard interactions would also change the physics of neutrino decoupling, such that a complete study would require solving the neutrino transport equations (\emph{à la} \cite{Grohs:2015tfy,Akita:2020szl,Froustey:2020mcq,Bennett:2020zkv,Froustey:2022sla}) including these non-standard processes.

Since BBN is only really sensitive to the neutrino distributions as local thermodynamic equilibrium gets broken, that is, slightly before and at freeze-out (temperature $T_\mathrm{FO}$), for our exploratory study it is not necessary to describe accurately the whole neutrino decoupling process, as long as the physics included in our BBN code is self-consistent. To that end, we will assume \emph{instantaneous neutrino decoupling}\footnote{More precisely, we also assume that this decoupling occurs well above $T \sim m_e$, such that the entropy of $e^+e^-$ is only transferred to photons — this is the “ultra-relativistic” approximation in the nomenclature of~\cite{Bennett:2019ewm}.} at a temperature higher than $T_\mathrm{FO}$, and that for the range of temperatures covered by our BBN code, neutrinos have distributions given by Eq.~\eqref{eq:normalization_g}.

Specifically, we need to distinguish the temperature of the electromagnetic plasma ($T_\gamma$), shared by photons, electrons, positrons and baryons, and the temperature $T_\nu$ neutrinos would have without distortions. In our approximation, neutrinos have decoupled such that $T_\nu \propto a^{-1}$, with $a$ the scale factor, while $T_\gamma$ is modified by $e^+e^-$ annihilations, which leads to a ratio $T_\gamma/T_\nu \simeq 1.40$ after annihilations. The distorted distribution $f_\nu(x,y_g)$ is defined with respect to this temperature $T_\nu$, such that $x=E/T_\nu$ in Eq.~\eqref{eq:normalization_g}, with $E$ the neutrino energy.

\subsection{Consequences on cosmological observables}
\label{subsec:consequence_cosmo}

\subsubsection{Energy density and $\Neff$ parameter}

After electron-positron annihilations (i.e., for $T_\gamma \ll m_e \simeq 0.511 \, \mathrm{MeV}$), the energy density of ultrarelativistic species can be written as
\begin{equation}
\label{eq:Neff}
    \rho_\mathrm{rad} = \rho_\gamma + \sum_{\alpha = e,\mu,\tau}{[\rho_{\nu_\alpha}+\rho_{\bar{\nu}_\alpha}]} + \rho_{\Delta N} \equiv \left[1 + \frac78 \left(\frac{4}{11}\right)^{4/3} \Neff \right] \frac{\pi^2}{15} T_\gamma^4 \, .
\end{equation}
We have explicitly written the total energy density $\rho_\mathrm{rad}$ as the sum of the photon energy density ($\rho_\gamma$), the sum of neutrino and antineutrino energy densities ($\sum_{\alpha=e,\mu,\tau}{[\rho_{\nu_\alpha}+\rho_{\bar{\nu}_\alpha}]}$), and the contribution from other, dark sector ultrarelativistic particles ($\rho_{\Delta N}$). First, we note that the energy density of (anti)neutrinos is modified when considering gray distributions, following Eq.~\eqref{eq:rho_yg}. 

Furthermore, we parametrize the additional dark radiation contribution with an extra number of degrees of freedom $\Delta N$, such that
\begin{equation}
\label{eq:rho_DeltaN}
    \rho_{\Delta N} = \frac78 \left(\frac{4}{11}\right)^{4/3} \Delta N \times \frac{\pi^2}{15} T_\gamma^4 \, .
\end{equation}
More specifically, we define $\Delta N$ such that the case $\Delta N = 0$ (with $y_g=0$ as well) corresponds to $\Neff = N_\mathrm{eff}^\mathrm{std} \simeq 3.010$. This value corresponds to the standard prediction in the instantaneous decoupling scenario (see discussion above), including QED corrections to the plasma thermodynamics\footnote{We emphasize that an “instantaneous decoupling” is not incompatible with $\Neff \neq 3$, since a value of exactly 3 requires to assume that the $e^+e^-$ were ultrarelativistic at the time of neutrino decoupling (which we assume), \emph{and} that the thermodynamics of the plasma do not include QED corrections (but we include them), see notably the Conclusions of Ref.~\cite{Bennett:2019ewm}.} (see, e.g., Table 1 in~\cite{Bennett:2019ewm} or Table I in~\cite{Froustey:2019owm}). Making this instantaneous decoupling approximation means that our $\Neff$ predictions may have an error of the order of 1 \%, since the “true” standard value is $N_\mathrm{eff}^\mathrm{std} = 3.044$~\cite{Froustey:2020mcq,Bennett:2020zkv,Drewes:2024wbw}. This error is still well below current experimental uncertainties on $\Neff$.
In our analysis, the observational constraint on $\Neff$, which is a constraint on $\rho_\mathrm{rad}$, allows us to constrain simultaneously the distortion parameter $y_g$ and $\Delta N$, since $\Neff^\mathrm{obs} = \Neff^\mathrm{std}(1 + y_g) + \Delta N$, thanks to Eqs.~\eqref{eq:rho_yg}, \eqref{eq:Neff} and~\eqref{eq:rho_DeltaN}.

Finally, we note that we will also allow the $\Delta N$ parameter to be negative. Even though this is incompatible with the interpretation of $\Delta N$ as additional dark radiation, there are a range of physical scenarios that can decrease $\Neff$, which would correspond to $\Delta N < 0$, see e.g.,~\cite{Steigman:2013yua, Chluba:2020oip, Sanchis:2025awq, Cadamuro:2011fd,Fuller:2011qy,Boyarsky:2021yoh,Rasmussen:2021kbf,Ovchynnikov:2024xyd,Akita:2024nam}. Current CMB experiments do not exclude this possibility, see for instance the recent ACT results~\cite{ACT:2025fju, ACT:2025tim}, which actually prefer $\Neff < 3$.

\begin{figure}[!ht]
    \centering
    \includegraphics[width=0.7\linewidth]{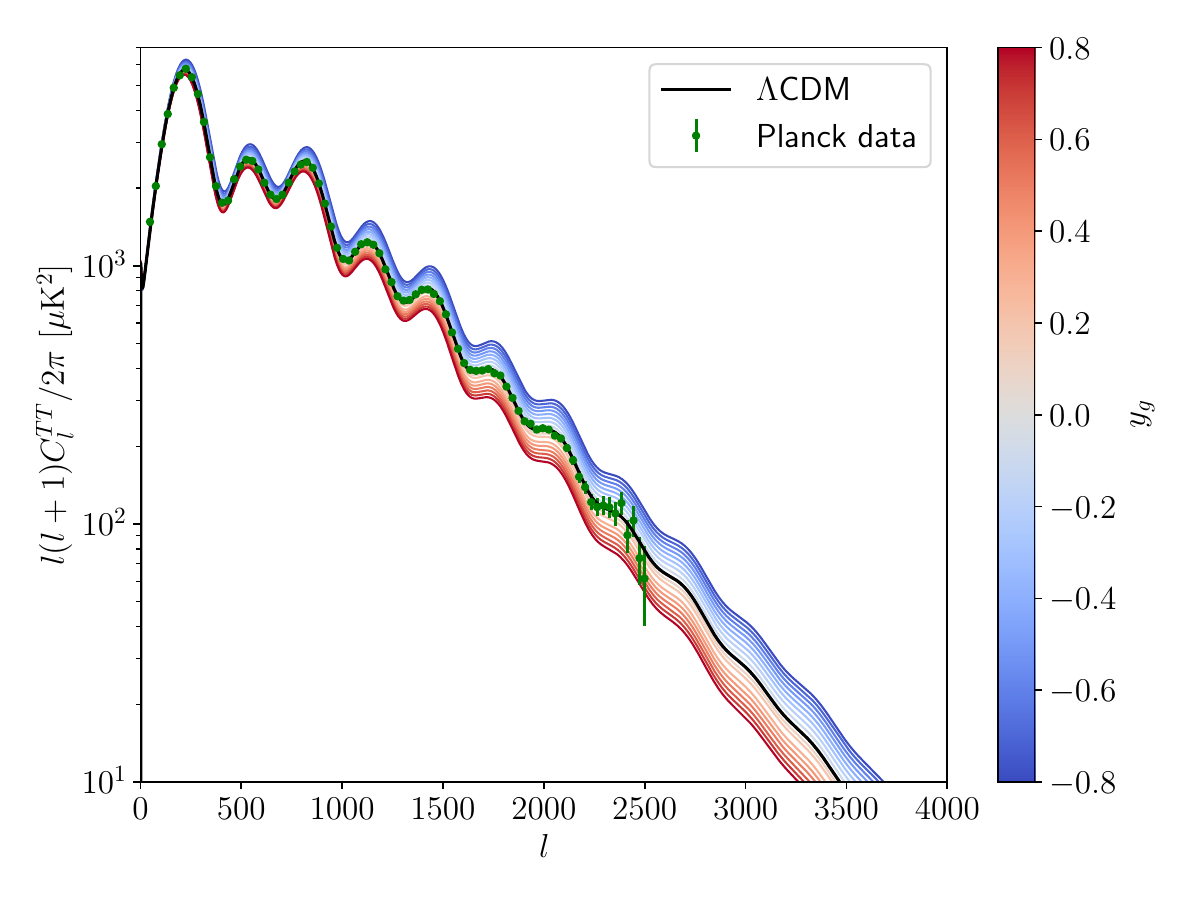}
    \caption{Effect of $y_g$ distortions on the CMB TT spectrum, assuming $\Delta N = 0$. The figure was generated using our modified version of CLASS~\cite{2011JCAP...07..034B_Class2}, see Sec.~\ref{ConstraintsFromObservations} for details.}
    \label{fig:ClTT_full}
\end{figure}

\begin{figure}[!ht]
    \centering
    \includegraphics[width=0.8\linewidth]{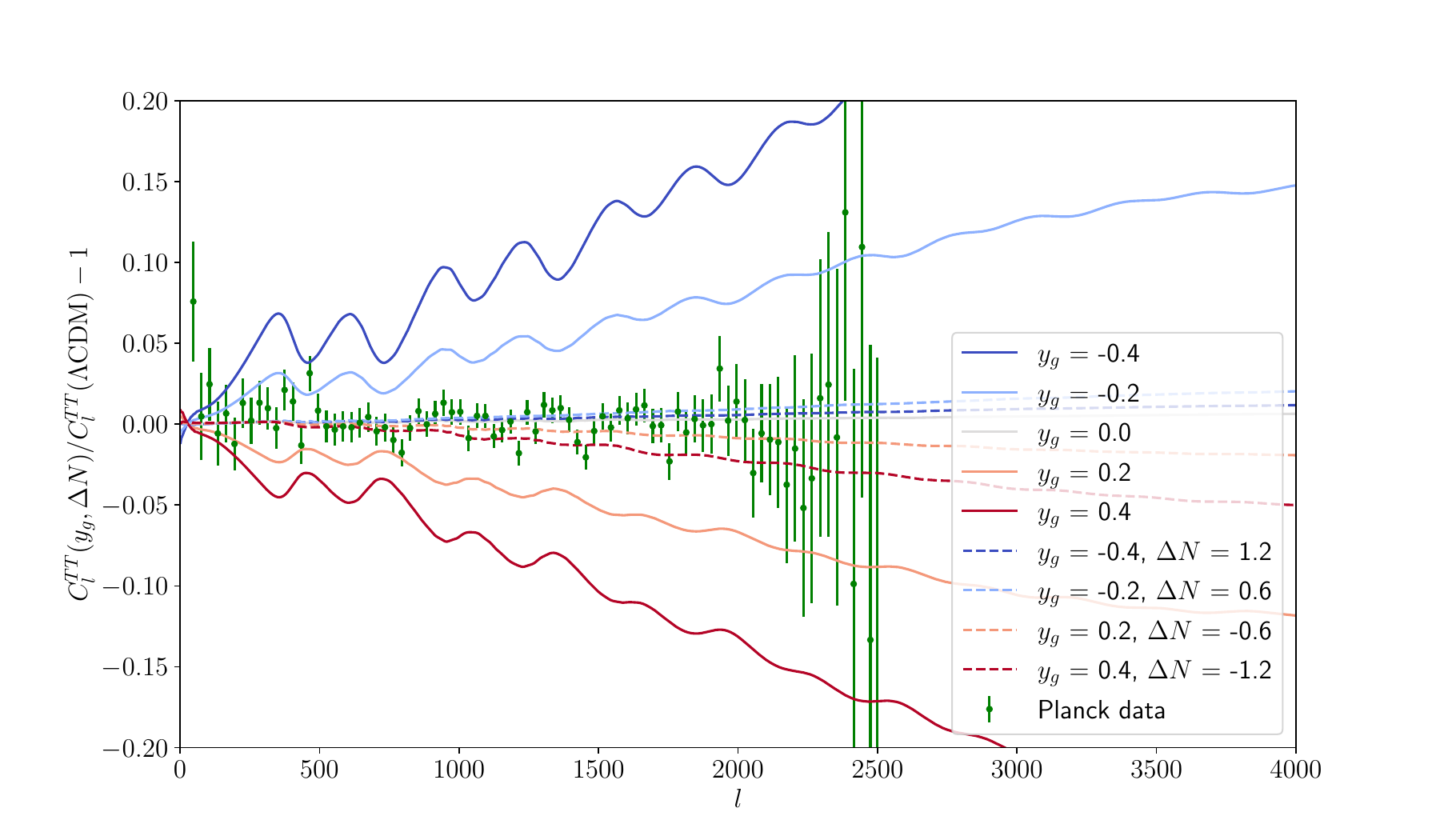}
    \caption{Relative change of the CMB TT spectrum with respect to $\Lambda \mathrm{CDM}$ in presence of $y_g$ distortions and $\Delta N$. We show that $\Delta N$ can roughly cancel the effect of $y_g$ and, even though the cancellation is not exact, the difference is too small given the current experimental sensitivity.}
    \label{fig:ClTT_cancel}
\end{figure}

The gravitational effects parametrized by $\Neff$ affect various cosmological observables, most notably the CMB. As an example, Fig.~\ref{fig:ClTT_full} shows the effect of varying $y_g$ (keeping $\Delta N = 0$ and all other cosmological parameters to their preferred values from Planck) on the CMB temperature-temperature angular power spectrum $C_l^{\mathrm{TT}}$, obtained with a modified version of the code CLASS~\cite{2011JCAP...07..034B_Class2} (see Sec.~\ref{ConstraintsFromObservations}). We also represent the data from Planck~\cite{Planck:2019nip}, which shows that non-zero values of $y_g$, for $\Delta N = 0$, would be excluded by measurements. However, if $\Delta N = - \Neff^\mathrm{std} y_g$, the combined effects cancel and $\Neff^\mathrm{obs} = \Neff^\mathrm{std}$. As the CMB's dependence on $y_g$ and $\Delta N$ comes mostly from this gravitational effect, this causes a degeneracy between the two parameters, as seen in Fig.~\ref{fig:ClTT_cancel}. The solid lines show the relative difference of $C_l^\mathrm{TT}$ compared to the unperturbed baseline, with $\Delta N = 0$. The dashed lines show, for the same value of $y_g$, the results with $\Delta N \simeq - \Neff^\mathrm{std} y_g$, with a systematically reduced deviation from standard $\Lambda$CDM compared to $\Delta N = 0$. Even though there are secondary effects that cause the cancellation not to be complete (in particular, the CMB damping tail has a mild dependence on the helium abundance, which, as we will see later, has different dependencies on $y_g$ and $\Delta N$), the non-cancellation is too small given current experimental bounds. We thus expect a significant degeneracy between $y_g$ and $\Delta N$ in our analysis from CMB measurements only (see Sec.~\ref{ConstraintsFromObservations}).

%%%%%%%%%%%%%%%%%%%%%%%%%%%%%%%%%%%%%%%
\subsubsection{Neutrinos and BBN}

Primordial nucleosynthesis is a key channel to constrain neutrino physics, as the conditions in which BBN takes place—and thus the primordial abundances—are directly dependent on (anti)neutrino distributions.

At the onset of BBN, almost all free neutrons are fused into helium-4, with small amounts of deuterium and traces of heavier elements (e.g., lithium). The final helium abundance is then directly set by the amount of free neutrons “available” at BBN. This number is usually quantified via the neutron-to-proton ratio $n_n/n_p$, which is set by the following weak reactions:
\begin{subequations}
\label{eq:weakreactions}
\begin{align}
    n + \nu_e &\longleftrightarrow p + e^- \, , \label{eq:weak_electron} \\
    n &\longleftrightarrow p + e^- + \bar{\nu}_e \, , \label{eq:neutron_decay} \\
    n + e^+ &\longleftrightarrow p + \bar{\nu}_e \, . \label{eq:weak_positron}
\end{align}
\end{subequations}
These reactions freeze out at a temperature $T_\mathrm{FO} \sim 0.8 \, \mathrm{MeV}$. Subsequently, $n_n/n_p$ decreases as the only reaction left is neutron beta decay~\eqref{eq:neutron_decay}, until light elements are produced (deuterium burning) below $T_\mathrm{Nuc} \sim 0.07 \, \mathrm{MeV}$. Because of the reactions~\eqref{eq:weakreactions}, distortions in the spectra of $\nu_e$ and $\bar{\nu}_e$ will change the value of $n_n/n_p$, and thus directly modify the primordial abundances. Similarly, an asymmetry between $\nu_e$ and $\bar{\nu}_e$ distributions also modifies $n_n/n_p$, allowing for constraints on a primordial lepton asymmetry at the BBN epoch (see e.g.,~\cite{Simha:2008mt,Shimon:2010ug,Castorina:2012md,Barenboim:2016shh,Oldengott:2017tzj,Escudero:2022okz,Froustey:2024mgf,Lattanzi:2024hnq}).

A second, subdominant, effect of neutrinos on BBN is gravitational. Indeed, a larger value of $\Neff$ leads to a higher expansion rate for a given photon temperature [see Eq.~\eqref{eq:Neff}]. In addition to increasing the freeze-out temperature of reactions~\eqref{eq:weakreactions}, it also leaves less time for neutron decay before $T_\mathrm{Nuc}$ (“clock effect” discussed in~\cite{Dodelson:1992km,Fields:1992zb}), resulting in a higher value of $n_n/n_p$ at the onset of BBN.

In this work, the distortions~\eqref{eq:normalization_g} have both a gravitational effect (see previous section) and a direct one on the weak rates. We include all this physics directly in a BBN code presented in the next section. We highlight once again that we make the assumption of instantaneous neutrino decoupling, and that the distortions formed quickly after neutrino decoupling. It is important that these distortions be formed before the freeze-out of reactions~\eqref{eq:weakreactions}; otherwise, their only effect would be through the change of energy density, which is completely degenerate with the additional parameter $\Delta N$. A more complete analysis, left for future work, would track neutrino distributions throughout the decoupling epoch, including the new processes leading to the distortions~\eqref{eq:normalization_g}, and use them as input in a BBN calculation, a proven method in other contexts~\cite{Mangano:2005cc,Grohs:2015tfy,Froustey:2020mcq,Froustey:2024mgf,Li:2024gzf}.

%%%%%%%%%%%%%%%%%%%%%%%%%%
\section{BBN calculations}
\label{BBNTheoryCalculations}

We include the modified neutrino distributions~\eqref{eq:normalization_g} in the BBN code \primat~\cite{Pitrou:2018cgg}, extending the method used in Ref.~\cite{Froustey:2019owm}. In that work, in addition to the change in energy density due to distortions, the weak interaction rates of reactions~\eqref{eq:weakreactions} were modified at the Born approximation level (i.e., in the infinite nucleon mass approximation). Given the large distortions considered here, we include the effect of spectral distortions consistently with the various corrections to the weak rates derived in~\cite{Pitrou:2018cgg}, namely, radiative and finite nucleon mass corrections.

\begin{figure}[!ht]
    \centering
    \includegraphics[width=0.95\textwidth]{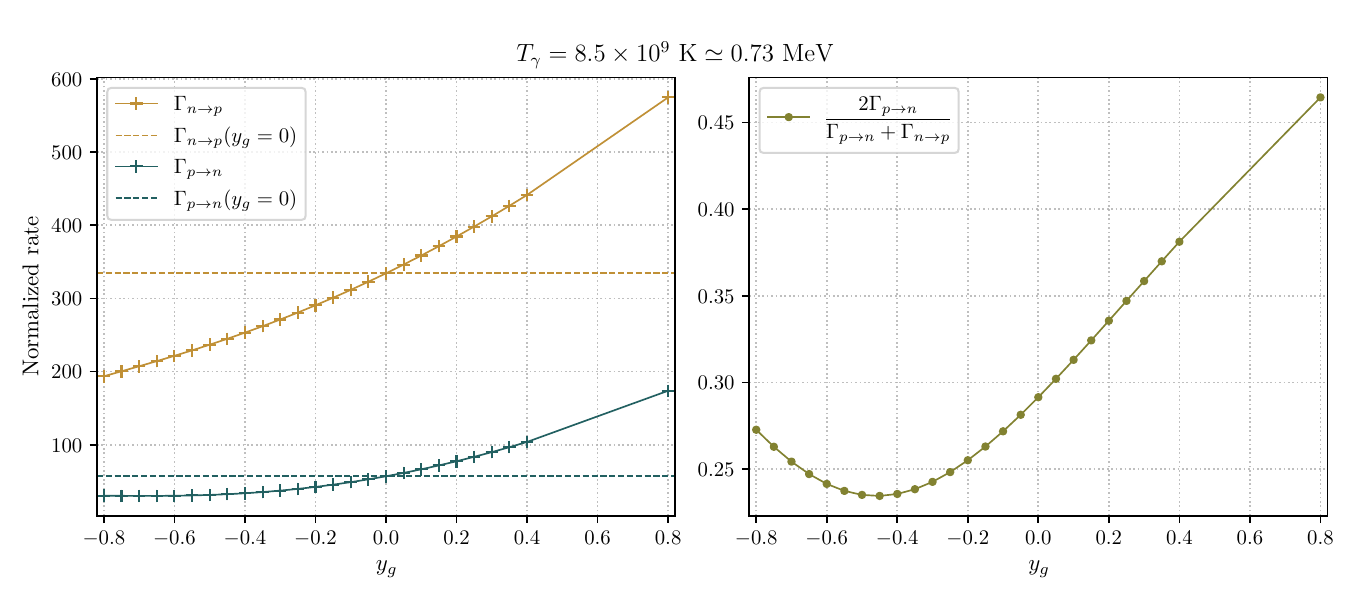}
    \caption{\emph{Left:} weak interaction rates, normalized by the neutron lifetime, as a function of the distortion parameter $y_g$ for a plasma temperature of $8.5 \times 10^9 \, \mathrm{K}$. The dashed lines are the values without distortions, used as a reference. \emph{Right:} combination of the rates providing a proxy for the helium-4 abundance [see Eq.~\eqref{eq:proxy_Yp} and surrounding text].}
    \label{fig:rates_comparison}
\end{figure}

To illustrate the change of neutron-to-proton interconversion rates with our modified neutrino distributions, we show on Fig.~\ref{fig:rates_comparison} the total rate of $n \to p$ reactions ($\Gamma_{n \to p}$, corresponding to Eq.~\eqref{eq:weakreactions}, from left to right) and of $p \to n$ reactions ($\Gamma_{p \to n}$, corresponding to Eq.~\eqref{eq:weakreactions}, from right to left). We choose a given temperature of $0.73 \, \mathrm{MeV}$, close to the freeze-out of $n \leftrightarrow p$ conversions in the standard case. Overall, a positive (resp. negative) $y_g$ corresponds to an increase (resp. decrease) of the rates, consistently with the shift of the neutrino distributions to higher (resp. lower) energies, see Fig.~\ref{fig:ydistortions}. The main difference between $n \to p$ and $p \to n$ reactions is the presence of an energy threshold for the latter (due to the difference of masses $m_n - m_p \simeq 1.3 \, \mathrm{MeV}$). Notably, for $y_g > 0$, the $\Gamma_{p \to n}$ rates increase relatively more than the $\Gamma_{n \to p}$ ones, since the presence of higher energy antineutrinos not only increases the cross section, but also helps overcoming the energy threshold.

On the right panel of Fig.~\ref{fig:rates_comparison}, we plot a quantity closely related to the final helium abundance. At equilibrium, detailed balance sets the neutron-to-proton ratio $(n_n/n_p)_\mathrm{eq} = \Gamma_{p \to n}/\Gamma_{n \to p}$. Therefore, the neutron fraction at weak freeze-out is
\begin{equation}
\label{eq:proxy_Yp}
    \left.\frac{n_n}{n_n + n_p}\right\rvert_{T_\mathrm{FO}} = \left.\frac{\Gamma_{p \to n}}{\Gamma_{p \to n} + \Gamma_{n \to p}}\right\rvert_{T_\mathrm{FO}} \, .
\end{equation}
Because of neutron decay, this ratio decreases between $T_\mathrm{FO}$ and deuterium burning at $T_\mathrm{Nuc}$, and the final helium abundance is well described by $\Yp \sim 2 n_n/(n_n+n_p)\rvert_{T_\mathrm{Nuc}}$ (see, e.g.,~\cite{Froustey:2019owm} for more details). That is why we plot the quantity of the right-hand side of Eq.~\eqref{eq:proxy_Yp}, with two caveats: first, the freeze-out temperature (for which, broadly speaking, $\Gamma_{n \leftrightarrow p}/H \sim 1$ with $H$ the Hubble rate) changes with $y_g$, but we plot the ratio of rates at a fixed value of $T_\gamma$; second, one would need to consider neutron decay to precisely connect this quantity to $\Yp$, and this phase is also slightly affected by the value of $y_g$. Regardless, this gives a good idea of the connection between the modified rates and what we can expect for the helium-4 abundance.

\begin{figure}[!ht]
    \centering
    \includegraphics[width=\textwidth]{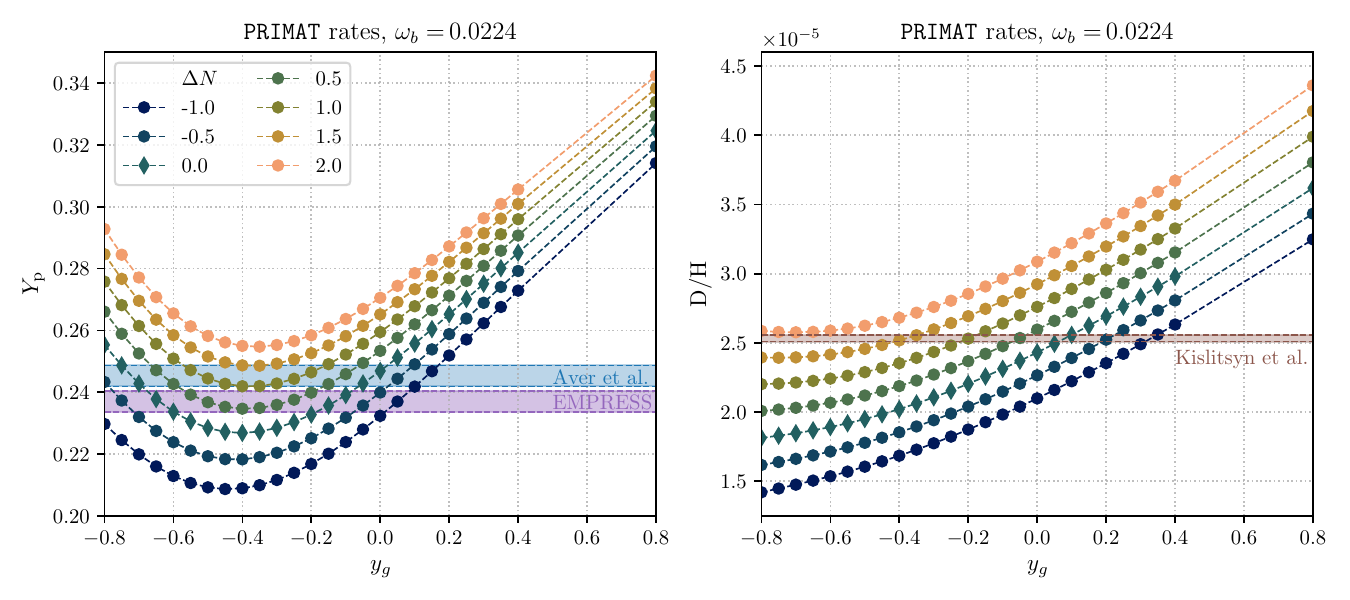}
    \caption{Final helium-4 (left) and deuterium (right) abundances obtained from \primat for different values of the distortion parameter $y_g$ and the generalized extra dark radiation contribution $\Delta N$. We identify with colored bands the measured helium abundances from Aver et al.~\cite{Aver2020} and EMPRESS~\cite{Matsumoto:2022tlr} (see Sec.~\ref{subsec:Empress}), and the measured deuterium abundance from Kislitsyn et al.~\cite{Kislitsyn:2024jvk}. The mention of “\primat rates” is in opposition to the choice of nuclear rates made in the BBN code \parthenope, see Sec.~\ref{subsec:parthenope}.}
    \label{fig:output_primat}
\end{figure}

This picture is confirmed in Fig.~\ref{fig:output_primat}, where we plot the \primat results for the helium-4 ($\Yp$) and deuterium ($\DH$) abundances, setting the baryon abundance value to $\Omega_b h^2 = 0.0224$ (preferred value from Planck~\cite{Planck:2018vyg}). We also show the experimental measurements with colored bands, see Sec.~\ref{ConstraintsFromObservations} for details.

%%%%%%%%%%%%%%%%%%%%%%%%%%%%%
\section{Constraints from observations}
\label{ConstraintsFromObservations}
%%%%%%%%%%%%%%%%%%%%%%%%%%%%%

Given their effects on cosmological observables (see Sec.~\ref{subsec:consequence_cosmo}), the two parameters $\{y_g, \Delta N\}$ can be constrained with observations. Here, we will use observations of primordial abundances from BBN and constraints on the cosmological $N_\mathrm{eff}$ parameter coming from non-BBN observations, namely the cosmic microwave background, CMB lensing and baryon acoustic oscillations (BAO). More specifically, we use CMB and CMB lensing observations from Planck~\cite{Planck:2019nip, Planck:2018lbu}, and BAO observations from DESI~\cite{DESI:2025zgx}.

For the observed BBN abundances, we employ the reference value for the helium-4 abundance by Aver et al.~\cite{Aver2020}, the deuterium abundance value from Kislitsyn et al.~\cite{Kislitsyn:2024jvk} (see also~\cite{Cooke:2017cwo,Guarneri:2024qxi}): 
\begin{align}
    \Yp^\obs &= 0.2453 \pm 0.0034 &&\text{\protect\cite{Aver2020}} \, , \label{eq:Yp_mes} \\
    %\DH^\obs &= (2.527 \pm 0.030) \times 10^{-5} %&&\cite{Cooke:2017cwo} \, . \label{eq:DH_mes}
    \DH^\obs &= (2.533 \pm 0.024) \times 10^{-5} &&\text{\protect\cite{Kislitsyn:2024jvk}} \, .\label{eq:DH_mes} %\\
    %\mathrm{Li/H}^\obs &= (1.6 \pm 0.3) \times 10^{-10} &&\text{\protect\cite{ParticleDataGroup:2024cfk}} \, .\label{eq:LiH_mes}
\end{align}
We report these values as colored bands on Fig.~\ref{fig:output_primat}.

One could also consider the lithium-7 abundance, whose recommended value by the Particle Data Group~\cite{ParticleDataGroup:2024cfk} (using observational data from~\cite{Sbordone_lithium_2010}) is $\mathrm{Li/H}^\obs = (1.6 \pm 0.3) \times 10^{-10}$. The lithium observations are known to be in strong tension with predictions from standard BBN theory—the so-called “lithium problem”~\cite{Fields:2011zzb}. Although one could hope that a combination of our non-standard parameters could satisfy the observations~\eqref{eq:Yp_mes}--\eqref{eq:DH_mes} and provide a lithium abundance compatible with observations, this is not the case. The parameter values required for the observed lithium abundance are in large tension with the parameter space preferred for helium and deuterium measurements. Given the uncertain status of the lithium problem~\cite{ParticleDataGroup:2024cfk}, we leave the lithium abundance out of the BBN observables used for our constraints.

To calculate the parameter constraints, we employ the Markov chain Monte Carlo (MCMC) code Cobaya~\cite{Torrado:2020dgo} and its MCMC sampler~\cite{Lewis:2002ah_mcmc, Lewis:2013hha_mcmc, Neal:2005_mcmc_dragging}. 
The inference of cosmological parameters from observations in Cobaya requires likelihoods associated to BBN and CMB + lensing + BAO measurements. For BBN observations, we employ Gaussian likelihoods around the values of the measured abundances. To calculate the predicted abundances for a given set of parameters ($y_g$, $\Delta N$ and baryon abundance $\omega_b \equiv \Omega_b h^2$), we construct a table with the code \primat (see Sec.~\ref{BBNTheoryCalculations}).
To get the standard deviation of the Gaussian likelihoods, we quadratically add the observational error and the error of \primat's predictions due to nuclear rates uncertainties. We use the code CLASS~\cite{2011JCAP...07..034B_Class2} to calculate the other cosmological observables (CMB, lensing, BAO). There are eight sampled parameters: the six standard $\Lambda$CDM cosmological parameters, along with $\Delta N$ and $y_g$; $\Yp$ is not left to vary but is obtained from the \primat table. We thus have to modify CLASS at two levels: first, to include the effect of the distorted distributions~\eqref{eq:normalization_g} (instead of the pure Fermi-Dirac ones) \footnote{In the results that we show, to be able to consider the case $\Delta N < 0$, we implement the distorsions in CLASS in an effective way that only considers the effect on the energy density and the helium fraction, setting $\Neff$ and $\Yp$ to the corresponding value given $y_g$ and $\Delta N$. In the regime $\Delta N \ge 0$, we have checked that this is consistent with what we get if we implement the full physical distribution function in CLASS.}, then, to read our BBN table, which has one more parameter than CLASS standard BBN tables because of the presence of the $y_g$ parameter.
We then use the CamSpec~\cite{2022MNRAS.517.4620R_CamSpec} likelihood, which makes use of Planck's PR4 (NPIPE)~\cite{Planck:2020olo_NPIPE} maps.

In principle, we can also constrain the baryon abundance, $\omega_b = \Omega_b h^2$, using BBN. For BBN-only constraints, we use a prior on this abundance obtained with a run including the aforementioned CMB, CMB lensing and BAO observations, but without using any BBN information. More specifically, this is a Gaussian prior with $\omega_b = 0.0223979 \pm 0.0001675$. In practice, we find that these observations are much more constraining on $\omega_b$ than BBN, so the $\omega_b$ constraints from the prior remain essentially unaltered.

In our model, we neglect the effect of neutrino masses, which we assume to be zero. In principle, one could also consider different non-zero neutrino masses to see how our model affects cosmological constraints on the neutrino mass. If we do this, we find that the effect of our model on these constraints is very small. The reason for this is that, once neutrinos become non-relativistic, their abundance is determined by their number density, so, since neutrino masses become relevant once neutrinos become non-relativistic, they will mostly be sensitive to the neutrino number density. However, as stated earlier, the neutrino distortion that we are considering conserves neutrino number density. The effect of the spectral distortions for constant number density is thus subdominant for that matter. We mention that a different distortion that does not conserve the number of particles could strongly affect neutrino mass bounds, but we do not study such a distortion here.

\subsection{Baseline results}

\begin{figure}[!ht]
    \centering
    \includegraphics[width=0.7\linewidth]{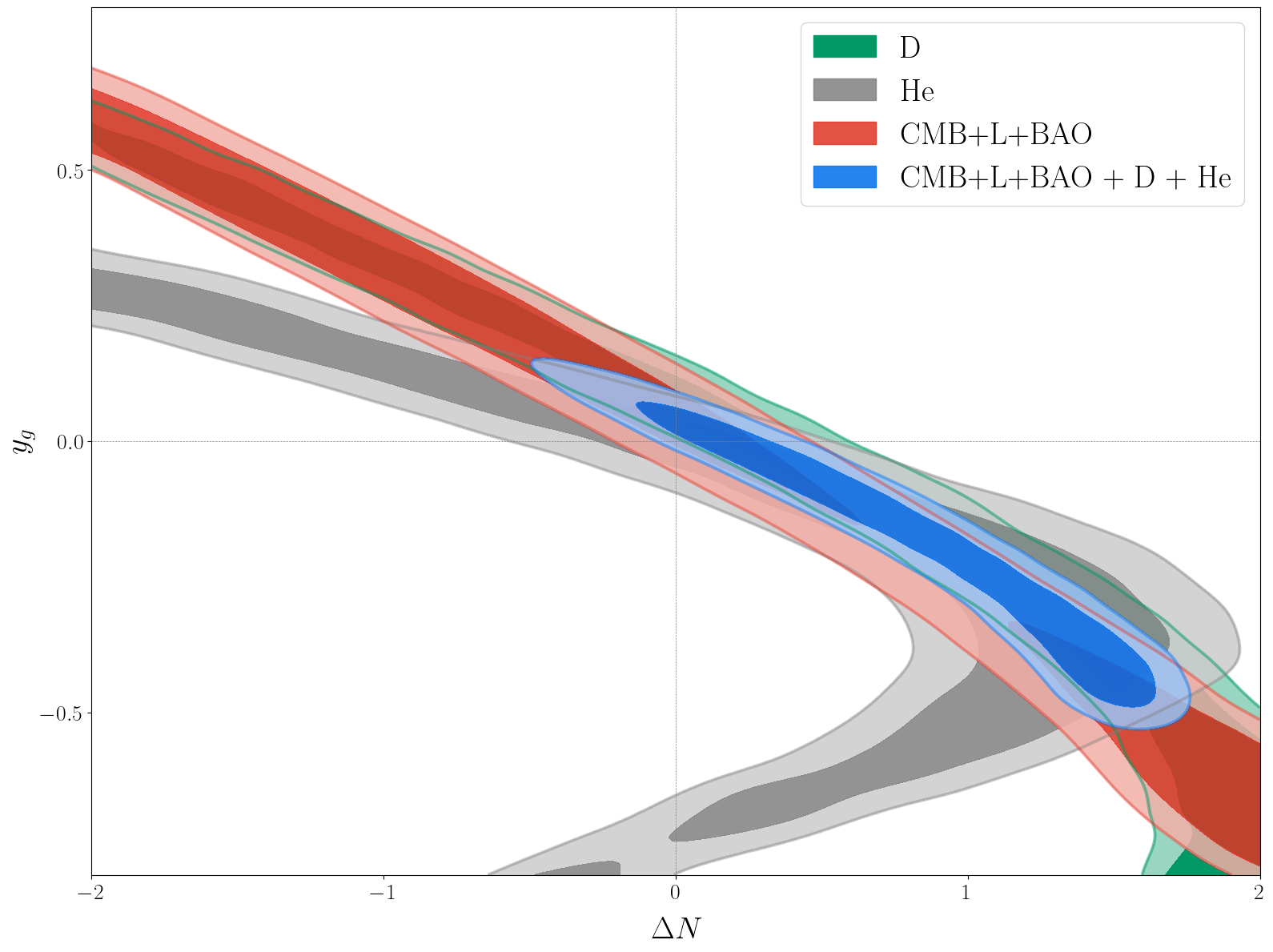}
    \caption{Constraints on additional relativistic degrees of freedom $\Delta N$ and the neutrino distortion parameter $y_g$ from deuterium observations (“D”, in green), helium-4 observations (“He”, in gray) and CMB + CMB lensing + BAO constraints (“CMB+L+BAO”, in red), with the combined constraints in blue. All constraints take into account BBN theory. Note that to obtain the CMB+L+BAO constraint, $\Yp$ is not a free parameter, but a derived parameter that is calculated from the base parameters using our BBN table.}
    \label{Compared_constraints}
\end{figure}

Figure~\ref{Compared_constraints} shows the constraints in the $(y_g,\Delta N)$ plane (marginalized over $\omega_b$) coming from the aforementioned observations: deuterium, helium, and CMB + lensing + BAO. The figure shows all 3 constraints separately, as well as the joint constraint that takes them all into account. Note that all three constraints contain a certain degeneracy between $\Delta N$ and $y_g$ that they cannot individually break. The main degeneracy is due to the equivalent gravitational effects of the $y_g$-distortion and some extra radiation, as discussed in Fig.~\ref{fig:ClTT_cancel}.
This leads to the “diagonal” degeneracy on Fig.~\ref{Compared_constraints}, which is notably broken by the strongly nonlinear dependence of the helium abundance on $y_g$ (see Fig.~\ref{fig:output_primat}). This dependence gives rise to the curved shape of the helium constraint in Fig.~\ref{Compared_constraints}. The slight sensitivity to $\Yp$ of CMB measurements also explains why the red contours on Fig.~\ref{Compared_constraints} are not perfectly along the $\Delta N \simeq - 3 y_g$ line.

The combined constraints can be used to put bounds on the $\{y_g, \Delta N\}$ parameters. Specifically, we can place the following bounds, where we quote errors at the 95\% confidence level:
\begin{align}
    y_g &= -0.20^{+0.29}_{-0.29} \, , \label{eq:yg_DeltaN_bounds} \\
    \Delta N &= 0.84^{+0.85}_{-1.01} \, .
\end{align}
Note that the bounds on a positive $y_g$ or a negative $\Delta N$ are very stringent, so those possibilities are highly disfavored, even if the change in the energy density is compensated by the other parameter.

For the possibility of a negative $y_g$ and a positive $\Delta N$, the bounds are more lenient, so, for example, an extra relativistic degree of freedom $\Delta N = 1$ could be allowed if a negative neutrino spectral distortion compensates the change in the energy density. However, we can rule out $\Delta N \ge 2$ at a confidence level over 99\%, regardless of whether $y_g$ compensates the energy density or not. At the time of BBN, taking into account the relation between the energy density and $y_g$ from Eq.~\eqref{eq:rho_yg}, we can rule out a scenario where the neutrino energy density is reduced by more than 50 \% via a gray distortion, even though this energy density is replaced by dark radiation.

%%%%%%%%%%%%%%%%%%%%%
\subsection{Alternative nuclear rates}
\label{subsec:parthenope}

\begin{figure}[!ht]
    \centering
    \includegraphics[width=\textwidth]{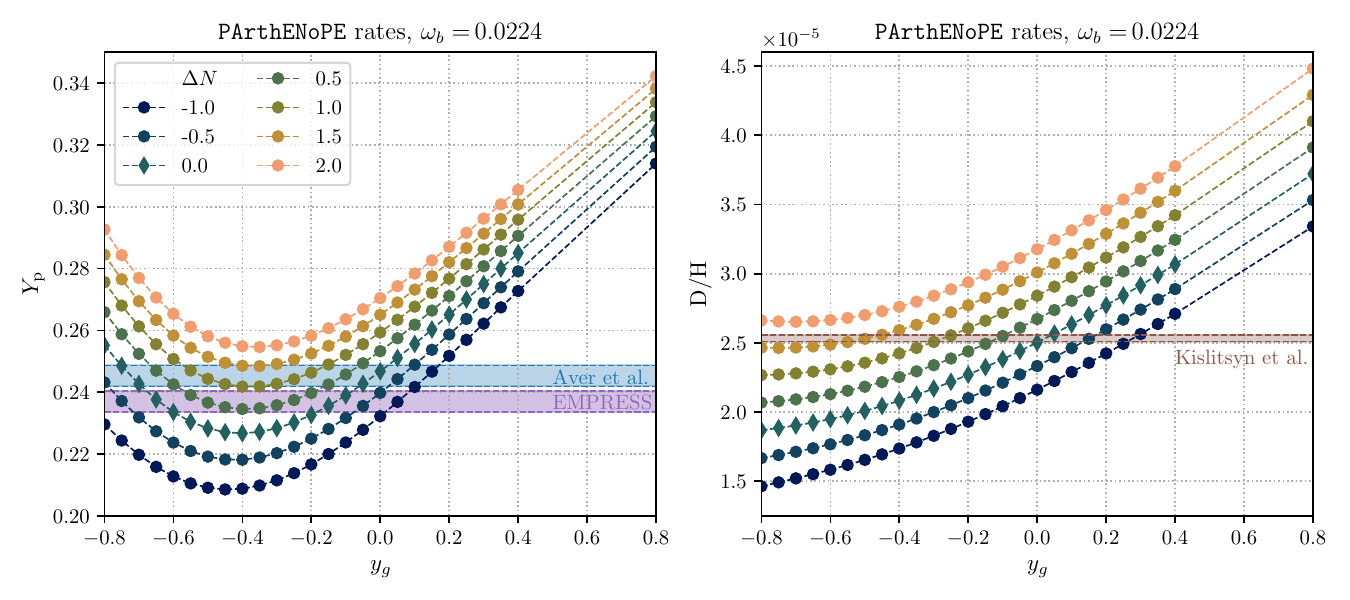}
    \caption{Final helium-4 (\emph{left}) and deuterium (\emph{right}) abundances obtained from \primat, modified to use the \parthenope rates for deuterium burning. Plotting conventions are identical to Fig.~\ref{fig:output_primat}.}
    \label{fig:output_parthenope}
\end{figure}

In standard cosmology, it is known that there is a mild discrepancy between \primat's predictions for the deuterium abundance and the actual measured abundance~\cite{Pitrou:2020etk}, as can be seen on the right panel of Fig.~\ref{fig:output_primat}, where the predicted abundance for $(y_g,\Delta N)=(0,0)$ is outside the range of Eq.~\eqref{eq:DH_mes}. This tension, which is absent from the predictions of the BBN code \parthenope~\cite{Gariazzo:2021iiu}, can be traced to different choices made in \primat and \parthenope regarding the rates of the nuclear reactions $\mathrm{D}(d,n){}^3\mathrm{He}$ and $\mathrm{D}(d,p){}^3\mathrm{H}$~\cite{Pisanti:2020efz,Pitrou:2021vqr}. 
In order to assess the robustness of our results, we have modified \primat in order to use the “\parthenope rates” for these deuterium burning reactions instead of the “\primat rates” (used in Fig.~\ref{fig:output_primat}), with the results for the abundances shown on Fig.~\ref{fig:output_parthenope}. 

Comparing Figs.~\ref{fig:output_primat} and \ref{fig:output_parthenope}, one can see, as expected, that the helium abundance is almost unchanged, while the deuterium predictions shift to higher values, reconciling the measurement~\eqref{eq:DH_mes} and the prediction for $\DH$ at $(y_g,\Delta N)=(0,0)$.

\begin{figure}[!ht]
    \centering
    \includegraphics[width=0.7\linewidth]{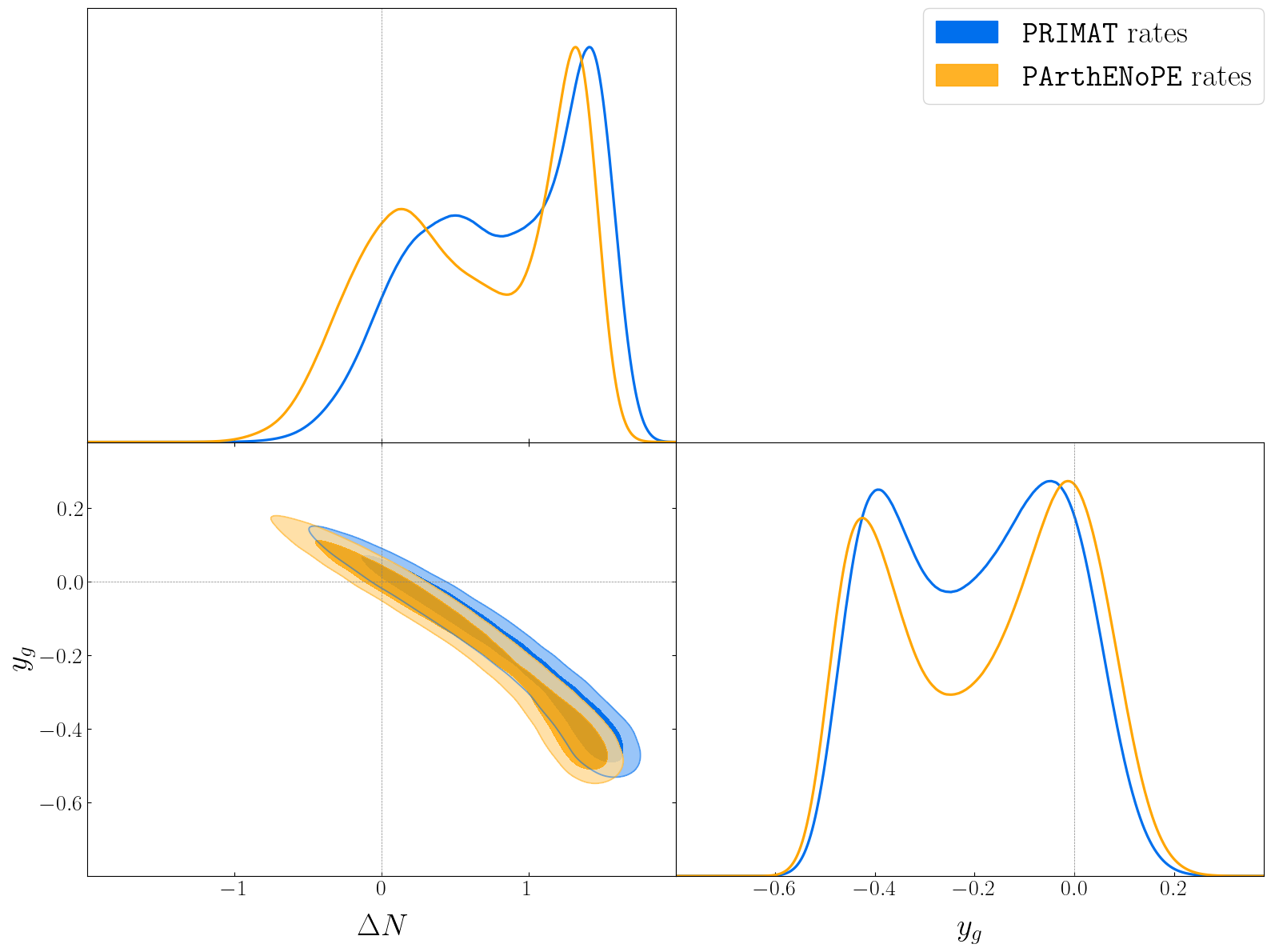}
    \caption{Combined constraints on additional relativistic degrees of freedom $\Delta N$ and the neutrino distortion parameter $y_g$ (see Fig.~\ref{Compared_constraints}), comparing constraints obtained with the \primat and \parthenope rates. Note that using the \parthenope rates increases the preference for the standard value, as the \primat rates are known to have a mild discrepancy with deuterium observations~\cite{Pitrou:2020etk}.}
    \label{Combined_constraints_Parthenope}
\end{figure}

The resulting combined constraints with these alternative abundance predictions can be seen in Fig.~\ref{Combined_constraints_Parthenope}. Overall, the preferred region is “shifted” towards the standard scenario, with the origin of the plot being just outside the $1\sigma$ region with the \primat rates (but still inside the $2\sigma$ limit), whereas it is inside the $1\sigma$ limit of the contour plot if one uses the \parthenope rates. This is consistent with the existence/absence of a mild deuterium tension for $(y_g,\Delta N)=(0,0)$ with \primat/\parthenope.

Figure~\ref{Combined_constraints_Parthenope} also shows the 1D posteriors for both parameters. Note that the 1D posteriors are clearly not Gaussian, and instead show two local maxima separated by a local minimum. Even though the minima are not deep enough to be significant, this shows that the distribution could potentially become bimodal with better data.

%%%%%%
\subsection{Alternative helium observations}
\label{subsec:Empress}
%%%%

Recently, the EMPRESS survey~\cite{Matsumoto:2022tlr} obtained a smaller value for $\Yp^\obs$, namely:
\begin{equation}
\label{eq:Yp_empress}
    \Yp^\obs\rvert_\mathrm{EMPRESS} = 0.2370^{+ 0.0034}_{-0.0033} \qquad \text{\protect\cite{Matsumoto:2022tlr}} \, .
\end{equation}
We report this value as a purple band on the left panels of Figs.~\ref{fig:output_primat} and \ref{fig:output_parthenope} ; it is $1.7\sigma$ below the traditional value, consistent with many observations and recommended by the Particle Data Group~\cite{ParticleDataGroup:2024cfk}. Such a smaller value can be accommodated by a nonzero lepton asymmetry in the electron neutrino sector at the epoch of BBN~\cite{Matsumoto:2022tlr,Burns:2022hkq,Escudero:2022okz,Froustey:2024mgf}, i.e., a neutrino $\mu$-distortion (see Sec.~\ref{SpectralDistortionsIntro}).
Here, we check in the following how our results change using the measurement \eqref{eq:Yp_empress} instead of \eqref{eq:Yp_mes}, exploring the potential of these generalized $y$-distortions to accommodate this helium abundance.

\begin{figure}[!ht]
    \centering
    \includegraphics[width=0.7\linewidth]{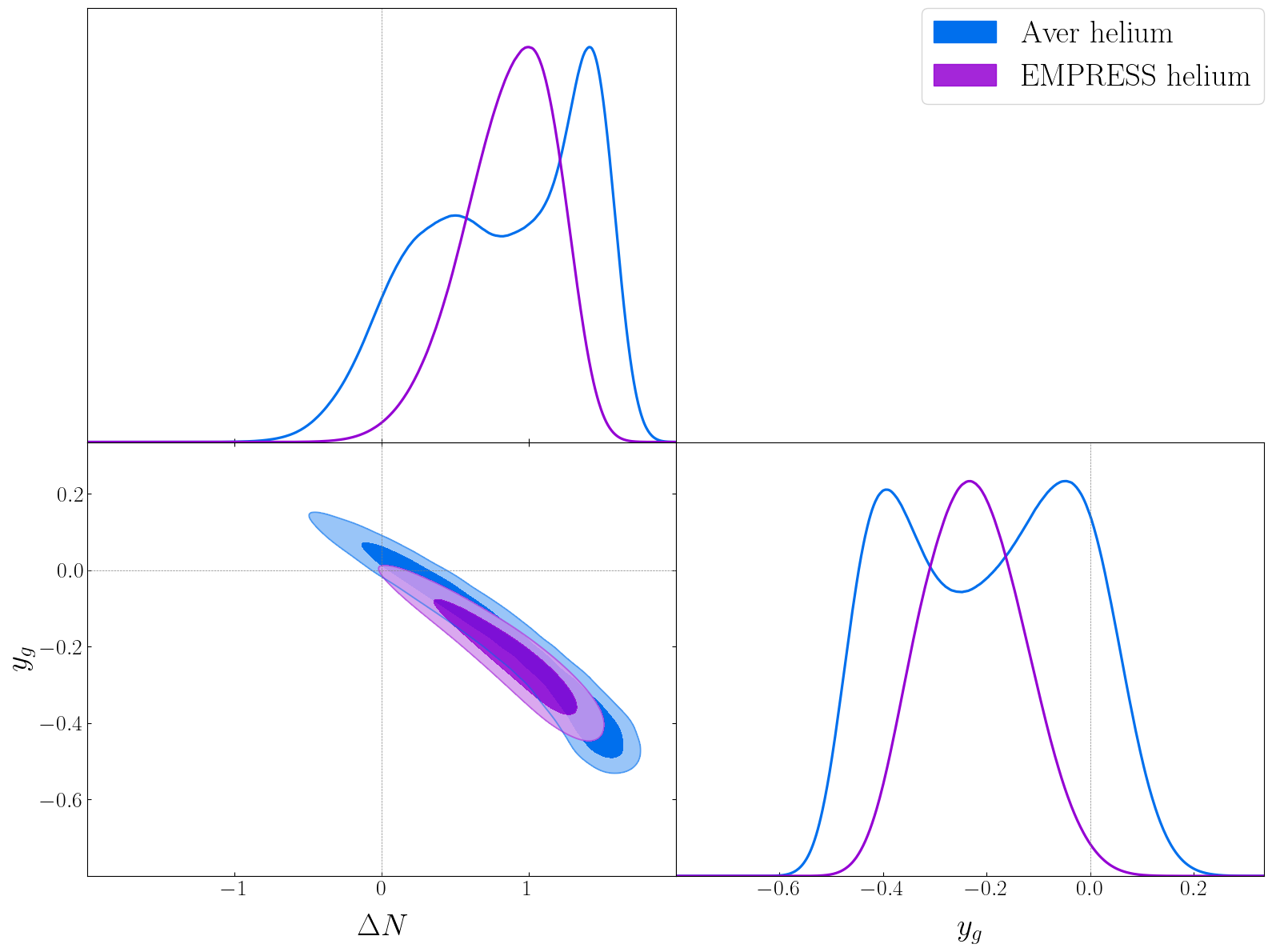}
    \caption{Combined constraints on additional relativistic degrees of freedom $\Delta N$ and the neutrino distortion parameter $y_g$ (see Fig.~\ref{Compared_constraints}), comparing constraints obtained with helium observations from Aver et al.~\cite{Aver2020} and from EMPRESS~\cite{Matsumoto:2022tlr}. Using the EMPRESS value, known to have a mild discrepancy with standard predictions, shows a preference for non-standard parameters $(y_g,\Delta N) \neq (0,0)$ at the $2 \sigma$ level.}
    \label{Combined_constraints_Empress}
\end{figure}

In Fig.~\ref{Combined_constraints_Empress}, we compare the combined constraints that we get using the standard value for the observed helium abundance, or using the EMPRESS value. In both cases, the \primat rates are used. The constraints using the EMPRESS measurement appear to be tighter, despite the EMPRESS helium observations~\eqref{eq:Yp_empress} having the same error as the reference value~\eqref{eq:Yp_mes}. This feature is due to the EMPRESS helium constraint having a smaller overlap with deuterium and CMB observations than the standard helium constraint. As in all the other cases considered in this work, there is an overall preference for a positive $\Delta N$ and negative $y_g$, but it is more significant in this case, as it is almost exactly at the $2\sigma$ level. We highlight, however, that this preference should be taken cautiously, both because of its low significance and because of the phenomenological nature of the model we considered here. Nevertheless, this clearly shows that scenarios beyond the standard paradigm of neutrino evolution in the MeV age are far from excluded, and even slightly preferred.

\section{Discussion and Conclusions}
\label{sec:Conclusions}

We have studied a model where extra dark radiation and generalized neutrino $y$-type distortions are added on top of the standard cosmological model. Specifically, the neutrino distortions that we have considered belong to a general class of “gray” distributions, a possibility largely overlooked in the neutrino community. For the distortions to have the effect described here, they need to be generated in the relatively narrow window that starts at neutrino decoupling and ends at the freeze-out of proton-neutron reactions \eqref{eq:weakreactions}. Even though this is a phenomenological model and we propose no specific mechanism for the origin of these distortions, we emphasize that any sufficiently efficient interaction that conserves the number of particles could cause a similar distortion. Therefore, such a distortion could come, for example, from interactions between neutrinos and this extra dark radiation, or between neutrinos and dark matter,\footnote{Note that, if dark matter is cold, its abundance at such early times is expected to be negligible, making it unlikely to cause sizable distortions. However, if dark matter is warm and becomes relativistic at very early times, its abundance could be non-negligible, allowing it to generate such distortions.} or between neutrinos and an extended dark sector, etc.

We have not taken into account the transfer of entropy from $e^-e^+$ as they become non-relativistic, as the way it affects the neutrino spectrum could be affected by the physics causing the distortion. For a given scenario, one would need to compute neutrino decoupling correctly (with a neutrino transport code \emph{à la} \cite{Grohs:2015tfy,Akita:2020szl,Froustey:2020mcq,Bennett:2020zkv}). This is beyond the scope of this article and left for future work. We note however that, in the standard case, this results in percent-level changes to the distributions, far below the size of the distortions allowed in our study.

The dark radiation and the spectral distortions have a very similar gravitational effect, which causes a degeneracy between them that the CMB and later-time observables cannot break. Nevertheless, we have found that BBN observations (in particular, observations of the primordial helium abundance) can break this degeneracy, thanks to the neutrino weak interactions' effect on the neutron-to-proton ratio at freeze-out.

This has allowed us to place simultaneous constraints on the parameters characterizing the dark radiation and the distortions, $\Delta N$ and $y_g$. These bounds strongly constrain scenarios with a positive $y_g$ (shift of the distribution to higher energy neutrinos) and a negative $\Delta N$. In contrast, scenarios with a positive $\Delta N$ and negative $y_g$ are more weakly constrained, although we can rule out a scenario where more than $\sim 1/2$ of the neutrino energy density is replaced by dark radiation. Nevertheless, we find a slight preference for a smaller replacement when using helium observations from EMPRESS~\cite{Matsumoto:2022tlr}, known to be in slight tension with standard helium predictions. To a lesser extent, we also find a slight preference for non-standard physics when employing the default nuclear rates used in \primat, known to cause a slight discrepancy between the predicted deuterium abundance in the standard scenario and the observed deuterium abundance. This highlights the importance of getting more (and more precise) BBN observations, as well as better measurements of the nuclear rates, as BBN remains our key probe of non-standard neutrino features in the early Universe. More generally, this work shows that the commonly accepted scenario of BBN-era physics can be significantly challenged, while remaining in excellent agreement (or even improving the agreement) with the few hints and clues left by the early Universe in cosmological observables.

\section*{Acknowledgements}

GB and HS are supported by the Spanish grants  CIPROM/2021/054 (Generalitat Valenciana),  PID2023-151418NB-I00 funded by MCIU/AEI/10.13039/501100011033/, and by the European ITN project HIDDeN (H2020-MSCA-ITN-2019/860881-HIDDeN). HS is also supported by the
grant FPU23/00257, MCIU. JF is supported by the Network for Neutrinos, Nuclear Astrophysics and Symmetries (N3AS), through the National Science Foundation Physics Frontier Center Award No.~PHY-2020275. We used Matplotlib~\cite{Hunter:2007ouj}, NumPy~\cite{Harris:2020xlr}, SciPy~\cite{2020SciPy-NMeth}, GetDist~\cite{2019arXiv191013970L_GetDist} and batlow~\cite{batlow} for calculations and generating plots.

\bibliography{Paper.bib}
\end{document}